\documentclass[twocolumn]{aastex631}
\usepackage[T1]{fontenc}
\let\tablenum\relax
\usepackage{siunitx}
\newcommand{\Arg}{\text{Arg}}

\begin{document}

\title{Dynamics and Origins of the Near-Resonant Kepler Planets}
\shorttitle{Near-Resonant Origins and Dynamics}
\shortauthors{Goldberg \& Batygin}

\correspondingauthor{Max Goldberg}
\email{mg@astro.caltech.edu}

\author[0000-0003-3868-3663]{Max Goldberg}
\affiliation{Department of Astronomy, California Institute of Technology\\Pasadena, CA 91125, USA}

\author[0000-0002-7094-7908]{Konstantin Batygin}
\affiliation{Division of Geological and Planetary Sciences, California Institute of Technology\\Pasadena, CA 91125, USA}

\begin{abstract}
Short-period super-Earths and mini-Neptunes encircle more than $\sim50\%$ of Sun-like stars and are relatively amenable to direct observational characterization. Despite this, environments in which these planets accrete are difficult to probe directly. Nevertheless, pairs of planets that are close to orbital resonances provide a unique window into the inner regions of protoplanetary disks, as they preserve the conditions of their formation, as well as the early evolution of their orbital architectures. In this work, we present a novel approach toward quantifying transit timing variations within multi-planetary systems and examine the near-resonant dynamics of over 100 planet pairs detected by \textit{Kepler}. Using an integrable model for first-order resonances, we find a clear transition from libration to circulation of the resonant angle at a period ratio of $\approx 0.6\%$ wide of exact resonance. The orbital properties of these systems indicate that they systematically lie far away from the resonant forced equilibrium. Cumulatively our modeling indicates that while orbital architectures shaped by strong disk damping or tidal dissipation are inconsistent with observations, a scenario where stochastic stirring by turbulent eddies augments the dissipative effects of protoplanetary disks reproduces several features of the data.
\end{abstract}

\section{Introduction}
The architectures of multiplanet systems hold crucial clues to their formation and evolution. Conditions of the protoplanetary disk leave deep imprints on individual and statistical properties of these planets and systems \citep{Lee2002,Adams2008,Lee2013,Millholland2019a}. Although the process of planet formation itself remains difficult to observe directly, the galactic planetary census is now broad enough to employ as a population-level testing ground for models of planetary formation \citep{Mordasini2015}.

One distinct feature is mean motion resonances: although most exoplanets are not near resonance, the small fraction that \textit{are} resonant imply that dissipative processes play an important role in at least those systems \citep{Batygin2015}. This small degree of certainty in their formation makes it possible to build more robust models incorporating other physical processes. Notably, numerous physical effects have been invoked to explain the unexpected phenomenon in which adjacent exoplanets tend to lie slightly wide of, rather than exactly on, mean motion resonances \citep{Delisle2012,Fabrycky2014,Terquem2019,Choksi2020}.

From a distinct point of view, near-resonant planets offer a unique opportunity to measure planetary masses and orbital elements normally invisible to transiting exoplanet surveys. When the ratio of orbital periods of adjacent planets is close to a ratio of small integers, planet-planet interactions are coherent and become amplified. The slight changes to the Keplerian orbits manifest as deviations from exact periodicity in the arrival of transits \citep{Agol2005a}. Inverting the transit timing variation (TTV) signal to produce mass and orbital constraints is nontrivial and subject to several degeneracies \citep{Lithwick2012a,Hadden2016}. Nevertheless, TTVs have routinely provided useful constraints on mass and orbits of small planets not amenable to radial velocity observations \citep{Hadden2014,Agol2021}.

Together, the special properties of near-resonant systems offer a unique opportunity to take precise measurements and use those constraints to reliably study their origin. Recently, studies of a small subset of resonant giant planets have produced possible histories of migration and constraints on the behavior of the inner disk \citep{Hadden2020,Nesvorny2022}. By extending this type of analysis to small planets and a much larger sample, we can test whether a given physical process acting in the late stage of planet formation could conceivably reproduce the current sample.

In Section \ref{sec:res}, we review the basics of first-order mean motion resonance from a Hamiltonian perspective. In Section \ref{sec:ttv}, we summarize the key properties of TTVs and then analyze the resonant structure of the \textit{Kepler} TTV sample. Section 4 tests four general models of the formation of near-resonant systems against our new constraints. We identify possible biases and possible future work in Section 5, and summarize our results in Section 6.

\section{Mean Motion Resonance}
\label{sec:res}
We begin with a brief overview of the Hamiltonian formulation of a first-order mean motion resonance. To first order in planet masses and eccentricities, the Hamiltonian for a pair of planets near a $k$:$k-1$ resonance is
\begin{eqnarray}
    \mathcal{H} = -\frac{\mathcal{G}M_*m_1}{2a_1} - \frac{\mathcal{G}M_*m_2}{2a_2} - \frac{\mathcal{G}m_1m_2}{a_2} \nonumber\\
    \times[fe_1\cos(k\lambda_2-(k-1)\lambda_1-\varpi_1)\nonumber\\
    + ge_2\cos(k\lambda_2-(k-1)\lambda_1-\varpi_2)]
    \label{eq:H1}
\end{eqnarray}
where $m$, $a$, $e$, $\lambda$, and $\varpi$ are the planet mass, semi-major axis, eccentricity, mean longitude, and longitude of pericenter, respectively \citep{Batygin2015}. Additionally, $\mathcal{G}$ is the gravitational constant, $M_*$ is the mass of the central star, and $f$ and $g$ are order-unity constants that depend on the resonant index \citep[see e.g.][]{Deck2013}. The two angles that appear as arguments of the cosines are referred to as ``resonant angles,'' and we will refer to them as $\phi_1$ and $\phi_2$, respectively. After a series of rescalings and canonical transformations that identify three conserved quantities, Eq. \ref{eq:H1} can be reduced to an integrable Hamiltonian with just one degree-of-freedom,
\begin{equation}
    \tilde{\mathcal{H}} = 3(\delta+1)\tilde{\Psi} - \tilde{\Psi}^2 - 2\sqrt{2\tilde{\Psi}}\cos(\psi).
    \label{eq:Hred}
\end{equation}
Here, $\psi$ is the coordinate and $\tilde{\Psi}$ is its conjugate momentum. Specifically, 
\begin{equation}
    \psi = k\lambda_2 - (k-1)\lambda_1 - \hat{\varpi}
    \label{eq:psi}
\end{equation}
where $\hat{\varpi}$ is a generalized longitude of pericenter, most easily expressed using complex eccentricities \citep{Hadden2019}\footnote{One can also write this equation as $\tan{\psi}=\frac{f e_1 \sin{\phi_1}+g e_2 \sin{\phi_2}}{f e_1 \cos{\phi_1} + g e_2 \cos{\phi_2}}$ \citep{Laune2022}.}:
\begin{equation}
\hat{\varpi} = \Arg({f e_1 e^{\iota \varpi_1} + g e_2 e^{\iota \varpi_2}})
\label{eq:hatvarpi}
\end{equation}
We will refer to $\psi$ as the ``mixed resonant angle.''
Finally, $\delta$ is a dimensionless constant that determines the topology of the Hamiltonian. Its form as derived in \cite{Batygin2015} is unwieldy but can be simplified dramatically in the compact orbits approximation, i.e. $k\approx k-1$, $a_1/a_2 \rightarrow 1$, $-f\approx g \approx 0.8k$ \citep{Deck2015}. In that case, $\delta$ takes the form
\begin{equation}
    \delta \approx -1  + \frac{1}{3}\left(\sigma^2 - \frac{2\Delta}{3k}\right)\left[\frac{15kM_*}{4(m_1+m_2)}\right]^{2/3}
\end{equation}
where 
\begin{equation}
\sigma^2=e_1^2+e_2^2-2e_1 e_2 \cos(\varpi_2-\varpi_1)
\end{equation}
is a generalized eccentricity and $\Delta$ is the normalized distance to exact resonance defined in terms of the inner and outer periods $P_1$ and $P_2$ \citep[e.g.][]{Lithwick2012a},
\begin{equation}
    \Delta = \frac{P_2}{P_1}\frac{k-1}{k} - 1.
\end{equation}
Additionally, in this approximation the action takes the form
\begin{equation}
    \tilde{\Psi} \approx \frac{1}{2}\sigma^2\left[\frac{15kM_*}{4(m_1+m_2)}\right]^{2/3}
\end{equation}
and the generalized longitude of pericenter is
\begin{equation}
    \hat{\varpi} \approx \Arg({e_1 e^{\iota \varpi_1} - e_2 e^{\iota \varpi_2}}).
\end{equation}
We will use the unapproximated forms of $\delta$ and $\hat{\varpi}$ unless indicated otherwise.

Hamiltonian~\ref{eq:Hred} has one, two, or three equilibria, depending on $\delta$. In all cases, there is a stable equilibrium point at $\psi=\pi$; this is the sole equilibrium for $\delta<0$. At $\delta=0$ another equilibrium appears, and for $\delta>0$ it splits into a stable and unstable equilibrium, both of which have $\psi=0$. The dynamical regime of the system can be categorized according to these fixed points. One regime is ``libration,'' in which $\psi$ oscillates around $0$ or $\pi$ in a bounded interval smaller than $[0,2\pi]$. Inversely, ``circulation'' indicates that $\psi$ eventually takes on every value from $[0,2\pi]$. Libration and circulation of a resonant angle are sometimes taken to be equivalent to being ``in resonance'' and ``nonresonant,'' respectively. However, resonance is in fact only \textit{formally} defined when the Hamiltonian has a separatrix that divides resonant and nonresonant trajectories, which only occurs when $\delta>0$ \citep{Henrard1983,Delisle2012}. Correspondingly, true resonant dynamics only occur for libration of $\psi$ around $\pi$ when $\delta>0$.

\section{Transit Timing}
\label{sec:ttv}
\subsection{Overview of TTVs}
Transit timing variations (TTVs) have proven to be an especially powerful tool in investigating the dynamics of near-resonant planetary systems. TTVs have been widely used to characterize planetary systems beyond what is obtainable from strictly periodic transits alone and sample posterior distributions of planet masses, eccentricities, and other orbital elements \citep[][and references therein]{Agol2021}. In principle, one could use these samples to compute the posterior distributions of the resonant angles in Hamiltonian \ref{eq:H1} for each system. In practice, however, the distributions of $\phi_1$ and $\phi_2$ are typically very broad and uninformative.

The reason for this limitation is that there are fundamental degeneracies in inverting TTV data. For example, the approximate first-harmonic TTV amplitude derived by \cite{Hadden2016} for the inner planet in a near-resonant pair is
\begin{equation}
    \delta t_\mathcal{F} \sim \frac{m_2 P_1}{2\pi M_*|\Delta|}\cdot \max \left\{1,\frac{|\mathcal{Z}|}{|\Delta|} \right\}
    \label{eq:TTV_amp}
\end{equation}
where
\begin{equation}
    \mathcal{Z} = \frac{f e_1 e^{\iota \varpi_1} + g e_2 e^{\iota \varpi_2}}{\sqrt{f^2 + g^2}}
\end{equation}
is a combined complex eccentricity relevant to the dynamics.\footnote{Note that $|\mathcal{Z}|\approx\sigma/\sqrt{2}$ in the compact approximation.} Higher order TTV signals still depend on planet eccentricities only through the combination $\mathcal{Z}$ rather than the individual planet eccentricities \citep{Hadden2016}.\footnote{The exception is the 2:1 resonance, where the second-harmonic TTV can individually constrain eccentricities \citep{Hadden2016}.} As a result, while $\mathcal{Z}$ is often well-constrained, $\varpi_1$ and $\varpi_2$ are rarely independently measured and the libration or circulation behavior of $\phi_1$ and $\phi_2$ cannot be determined \citep[e.g.][]{Petigura2018,Petigura2020}. The analytical results of \cite{Hadden2016} only apply to near-resonant systems that are not actually in resonance. Nevertheless, \cite{Nesvorny2016} find that TTVs for planets in resonance also depend only on $\mathcal{Z}$. 

However, a precise measurement of $\mathcal{Z}$ \textit{does} allow for a direct measurement of $\hat{\varpi}$, because, from Eq. \ref{eq:hatvarpi}, $\hat{\varpi}=\Arg{\mathcal{Z}}$. Thus, the dynamically relevant mixed resonant angle $\psi$ can be calculated. Remarkably, while TTV data typically cannot determine the full planetary orbits, they can constrain the parameters of the mean motion resonance. The ability of TTV data to precisely measure $\psi$ has been used previously to study the resonant behavior in a few systems \citep{Hadden2017,Petit2020a}. Here, we apply it on a population level to study many near-resonant pairs in a uniform way. 

\subsection{Observed TTV Systems}
Our sample is the near-resonant pairs of planets from \textit{Kepler} studied by \cite{Hadden2017} and \cite{Jontof-Hutter2021}. Each work ran N-body models to fit transit times of \textit{Kepler} systems and derived the mass, orbital period, time of midtransit, and eccentricity vector of each planet (orbits are assumed to be planar). We use their Markov Chain Monte Carlo posterior distributions of those parameters. If a planet pair was fit by both works, we use the posteriors from \cite{Hadden2017}. We remove planet pairs near second-order resonance (which have a different Hamiltonian) and pairs with $|\Delta|>0.1$, leaving 105 unique planet pairs in the joint sample. Of these, 40 are near the 2:1 resonance, 37 are near the 3:2 resonance, and the rest are near higher-index first-order resonances. We use the default set of mass/eccentricity priors from \cite{Hadden2017} for the following analysis, but checked that the results are the same for their alternative ``high mass'' priors.

\begin{figure*}
    \centering
    \includegraphics[width=0.5\textwidth]{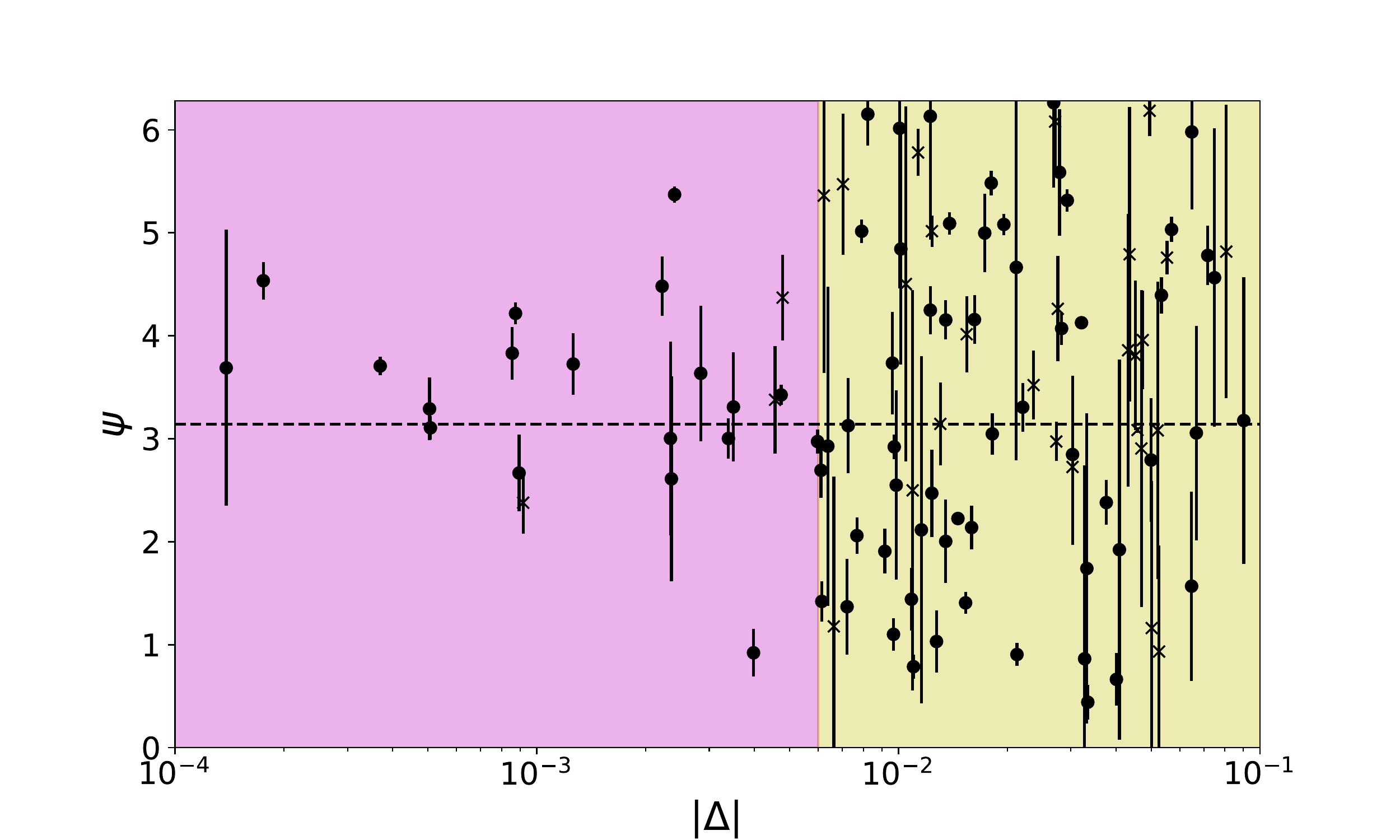}
    \includegraphics[width=0.4\textwidth]{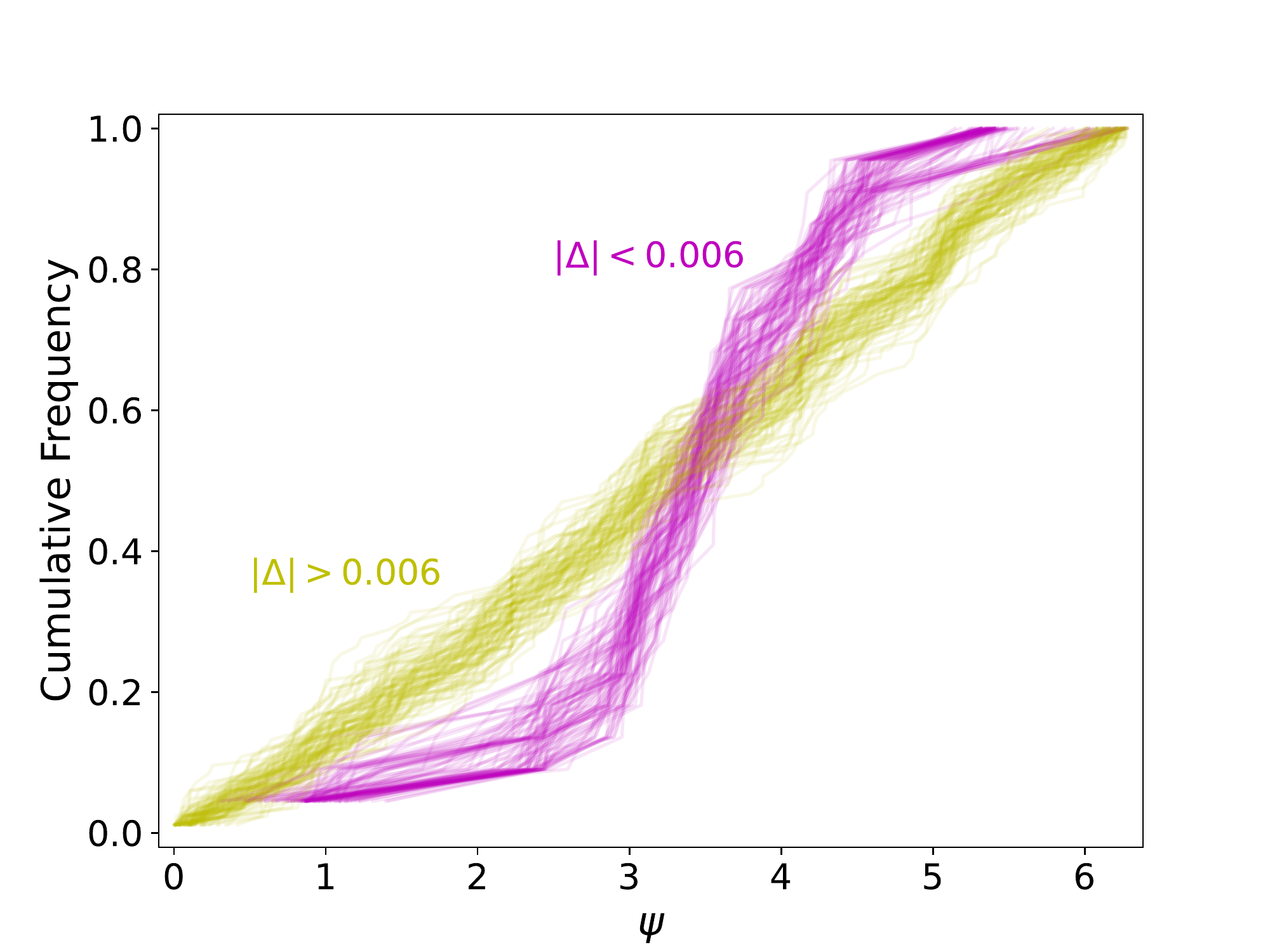}
    \caption{Left: Distance from exact resonance $\Delta$ versus the osculating mixed resonant angle $\psi$ for the 108 pairs of planets near first-order resonances characterized in \cite{Hadden2017} and \cite{Jontof-Hutter2021}. Dots mark pairs outside of exact resonance ($\Delta > 0$) and crosses mark pairs inside ($\Delta<0$). Error bars are $1\sigma$ circular standard deviations. The resonant equilibrium at $\psi=\pi$ is shown with a horizontal dashed line. Right: Cumulative distributions of $\psi$ for the systems nearest to resonance ($|\Delta|<0.006$, in maroon), and the more distant systems ($|\Delta|>0.006$, in yellow). In both cases, 100 CDFs are plotted by drawing from the posterior distribution of $\psi$ for each planet pair.}
    \label{fig:delta_psi}
\end{figure*}

For each of these planet pairs, we compute the osculating value of $\psi$ from the posterior distributions using Eqs. \ref{eq:psi} and \ref{eq:hatvarpi}. The results are shown in Figure \ref{fig:delta_psi}, where $\psi$ has been plotted against $|\Delta|$. Uncertainties on $\psi$ are computed using a circular standard deviation and are an average of $\sim 3$ times smaller than the uncertainties on $\phi_1$ and $\phi_2$. Figure \ref{fig:delta_psi} clearly shows two regimes: for $|\Delta|\lesssim 0.006$, $\psi$ is clustered around $\pi$. For $|\Delta|\gtrsim 0.006$, $\psi$ is uniformly distributed on $[0,2\pi]$.\footnote{The code for these and all following calculations is available at \url{https://github.com/goldbergmax/resonant-capture-simulation}.}

To confirm that these are indeed distinct distributions and not the result of a small sample size or measurement uncertainty, we randomly drew one value of $\psi$ from the posterior of each planet pair to construct two empirical cumulative distribution functions (CDFs) for pairs with $|\Delta|<0.006$ and $|\Delta|>0.006$. We then used the two-sample Kuiper test implemented by the \texttt{astropy} package to compute a false alarm probability (FAP) that these two samples were drawn from the same distribution. Repeating this process 10,000 times, we found that the median FAP was 0.0063. The right panel of Figure \ref{fig:delta_psi} is a representation of 100 of these empirical CDFs in each range of $\Delta$. Although the CDFs of the planet pairs with $|\Delta|>0.006$ seem to follow a uniform distribution, the pairs nearest to resonance show a clear kink in the distribution near $\pi$.

We noted in Section \ref{sec:res} that resonant trajectories are only formally defined if a separatrix exists, equivalent to $\delta>0$. Is this the case for the systems in Figure \ref{fig:delta_psi}? In the compact limit (see Section \ref{sec:res}) and assuming the system is at the stable equilibrium, the condition for the existence of a separatrix is
\begin{equation}
    \Delta < \Delta_\text{crit} \approx \frac32\left(\frac{4}{15}\frac{m_1+m_2}{M_*}\right)^{2/3} k^{1/3}.
    \label{eq:Deltacrit}
\end{equation}
As a typical near-resonant \textit{Kepler} system, we will assume $m_1\approx m_2\approx 10M_\oplus$, $M_*\approx M_\odot$, and $k\approx 3$. For this ``fiducial'' system, Eq. \ref{eq:Deltacrit} gives $\Delta_\text{crit}\approx~0.001$. This boundary is narrower than the break seen in Figure \ref{fig:delta_psi}, and indicates that the vast majority of our sample cannot strictly be resonant, regardless of $\psi$ \citep{Delisle2012}. Interesting, clustering persists beyond $\Delta_\text{crit}$ up to $\approx 0.006$, suggesting that libration exists in some formally nonresonant systems.

The resonant dynamics also induce a forced eccentricity on the pair of planets \citep{Lithwick2012a}. That is, the stable equilibrium of Hamiltonian \ref{eq:Hred} implies a nonzero value of $\sigma$ for $\Delta>0$. By solving the equilibrium equation for the Hamiltonian in the compact limit, one obtains
\begin{equation}
    \sigma_\text{eq} = \frac{4}{5\Delta}\frac{m_1+m_2}{M_*}.
    \label{eq:e_forc}
\end{equation}
A similar expression can be derived for the 2:1 resonance, and in that case the forced eccentricity is a factor of $\sim 2$ smaller than Eq. \ref{eq:e_forc} \citep{Lithwick2012a}.

\begin{figure}
    \centering
    \includegraphics[width=\columnwidth]{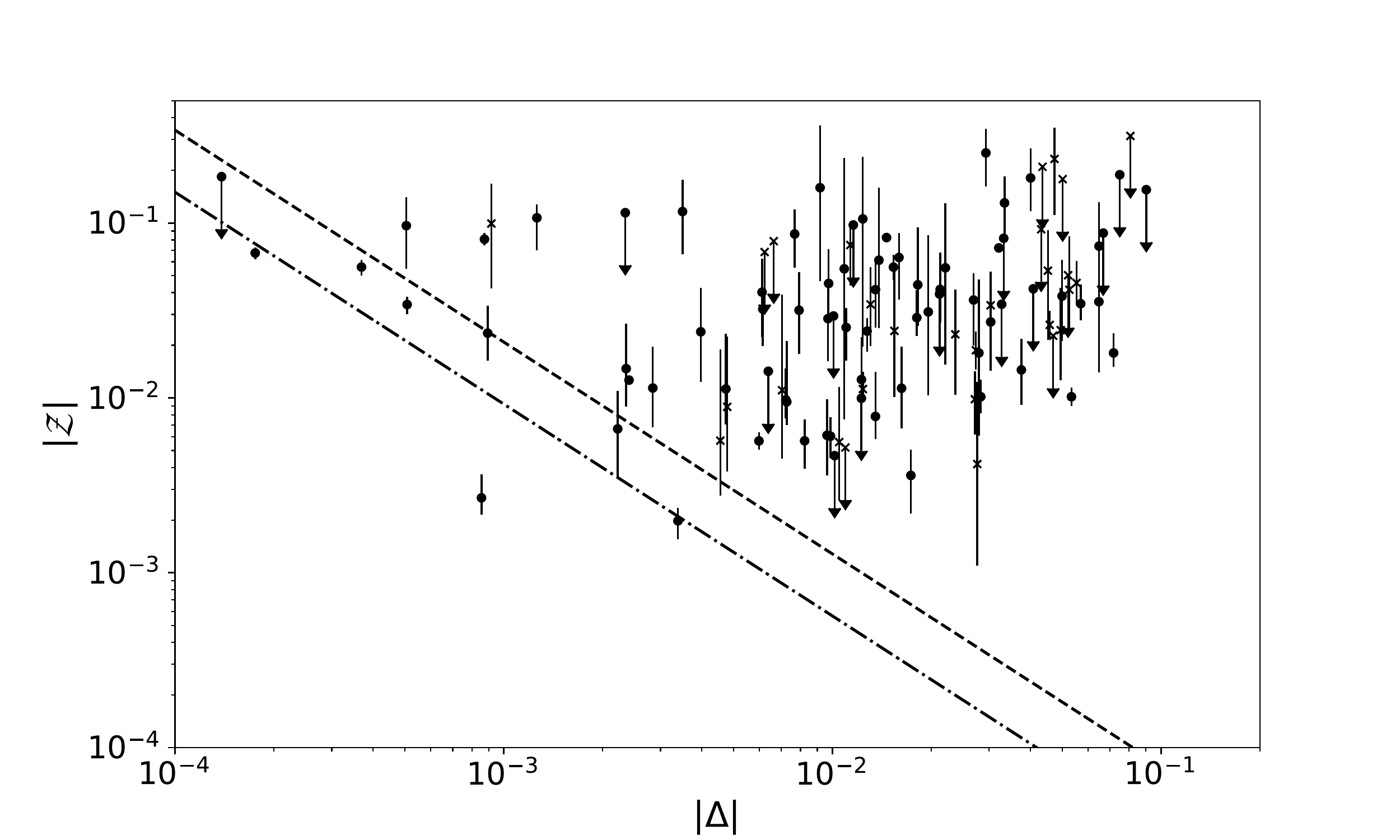}
    \caption{The norm of the generalized eccentricity $\mathcal{Z}$ as a function of distance from resonance $\Delta$ for the TTV systems. As in Figure \ref{fig:delta_psi}, circles and crosses mark pairs outside and inside of the resonance, respectively. For well-measured eccentricities, error bars are $1\sigma$ uncertainties, otherwise $2\sigma$ upper limits are plotted. The dashed and dash-dotted lines are the equilibrium forced eccentricities in the compact limit and for the 2:1 resonance, respectively (Eq. \ref{eq:e_forc}).}
    \label{fig:Delta_Z}
\end{figure}
Figure \ref{fig:Delta_Z} shows the observed $|\mathcal{Z}|$ for the TTV sample compared to the forced eccentricity $|\mathcal{Z}_\text{eq}|\approx \sigma_\text{eq}/\sqrt{2}$. Some eccentricities are not robustly measured and the positive definite nature of $|\mathcal{Z}|$ can introduce a bias. Following \cite{Hadden2017}, if the projection of the measured $\mathcal{Z}$ onto the median of its distribution is consistent with 0 at the $1\sigma$ level (true for about $25\%$ of systems), we report only upper limits on $|\mathcal{Z}|$. The observed eccentricities exceed the equilibrium forced value in most cases, often by an order of magnitude, except when $|\Delta| \lesssim 0.006$. In the context of Hamiltonian \ref{eq:Hred}, a free eccentricity exceeding the forced eccentricity implies that the $(\tilde{\Psi},\psi)$ phase space trajectory circumnavigates the origin and hence $\psi$ is in circulation. Random draws of $\psi$ from a circulating population will be distributed almost uniformly. Thus, the distribution of $\psi$ in Figure \ref{fig:delta_psi} is consistent with the observed $|\mathcal{Z}|$.

\subsection{Aside: Pitfalls of Determining Libration with Standard Resonant Angles}
\begin{figure}
    \centering
    \includegraphics[width=\columnwidth]{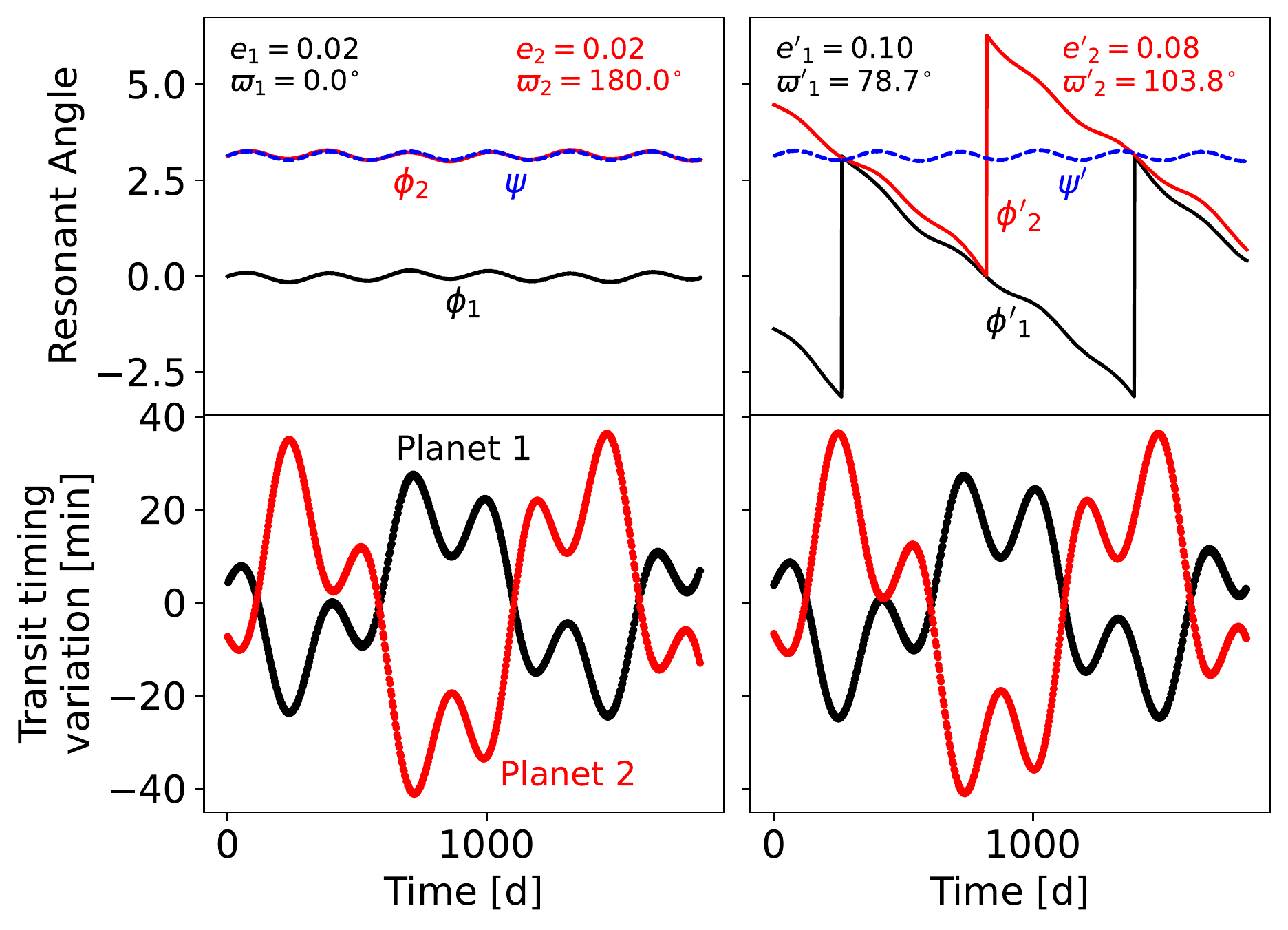}
    \caption{An example of the impact of two eccentricity configurations (left and right columns) on TTVs and inferred resonant angles. In the top row, the black and red lines are the standard resonant angles $\phi_1$ and $\phi_2$, respectively, while the dashed blue line is the mixed angle $\psi$. The bottom row shows the observed minus calculated transit times for the inner and outer planet in black and red. Without additional information, TTV fits cannot distinguish between the two cases and often recover solutions where $\phi_1$ and $\phi_2$ circulate even if the system is resonant.}
    \label{fig:ttv}
\end{figure}

While we have demonstrated that the mixed resonant angle $\psi$ is a valuable tool for studying near-resonant systems, it is worth emphasizing that considering only $\phi_1$ and $\phi_2$ as derived from TTV data can be deeply misleading. As an example, consider a coplanar pair of $10M_\oplus$ planets around a solar mass star with initial conditions $P_1=2$~d, $P_2=3$~d, $\lambda_1=\lambda_2=0$, and complex eccentricities $e_1=0.02\cdot e^{\iota 0}$, $e_2=0.02\cdot e^{\iota \pi}$. By numerically integrating this sytem with \texttt{rebound}, we find that the 3:2 resonant angles $\phi_1,\phi_2,\psi$ all librate with small amplitudes and the transit times vary by $\sim 30$~min over the resonant libration cycle (left column of Figure \ref{fig:ttv}).

Now suppose we construct a similar system (represented by primed coordinates) where we translate the complex eccentricities by some complex number $\xi$ so that $e'_1 = e_1 + \xi$ and $e'_2 = e_2 -(f/g)\xi$. All other orbital elements are left untouched. Hamiltonian \ref{eq:Hred} is invariant under such a transformation because $\tilde{\Psi}$ and $\psi$ depend on the eccentricities only through a linear combination in which $\xi$ cancels. Likewise, $|\mathcal{Z}|$ remains unchanged. The right panel of Figure \ref{fig:ttv} shows the evolution of the transformed system, where we have arbitrarily chosen $\xi=0.1\cdot e^{\iota \pi/2}$. As expected, $\psi'$ librates and the TTV signal is nearly identical. However, $\phi'_1$ and $\phi'_2$ are now in circulation, so an analysis which only computed these angles would classify this system as ``nonresonant.'' In general, TTV fits cannot distinguish between these two configurations \citep{Leleu2021a} and therefore they frequently recover both librating and circulating solutions to the same data, even if $\psi$ librates across nearly all of the allowed entire parameter space \citep{Dai2023}.

For completeness, we note that there are some ways to distinguish between the systems in the left and right columns of Figure \ref{fig:ttv}. Higher-order resonant terms and secular evolution depend on different combinations of the eccentricities, so higher-precision or longer baseline TTV measurements can break the degeneracy. The most eccentric solutions can also be eliminated via dynamical stability tests. Finally, transit durations can directly measure $e\sin{\varpi}$ for well-understood systems \citep{VanEylen2015}. Photodynamical analyses that incorporate all of these strategies can provide excellent constraints on resonant behavior, for example in K2-19 \citep{Petigura2020,Petit2020a}. The problem can also be avoided altogether in systems of three or more planets by considering zeroth-order Laplace angles that depend only weakly on eccentricity \citep{Siegel2021}.

\section{Migration History of Near-Resonant Systems}
Early on in the \textit{Kepler} mission, an excess of planet pairs just outside of first-order resonances was identified \citep{Lissauer2011,Fabrycky2014}. The effect is most striking near the 2:1 and 3:2 commensurabilities, where there is a distinct peak at $\Delta \sim 1\%$ and a trough inside the exact resonance \citep{Weiss2022}. There is considerable literature on this topic, and explanations of this unexpected feature have broadly fallen into two categories. 

The first idea argues that the departure from resonance occurs during the migration phase within the gaseous disk. As a pair of planets capture into resonance in the protoplanetary nebula, the orbital periods reach an equilibrium value of $\Delta$ that is set by the disk properties \citep{Terquem2019}. Various disk models and parametrizations have been found to match the distribution of period ratios wide of resonance \citep{Choksi2020}.

The second idea argues alternatively that the distancing from resonance occurs after the disappearance of the protoplanetary disk. Early theoretical studies noted that a near-resonant pair of planets, under some form of energy dissipation that conserves angular momentum, will ``repel,'' and the orbital period ratio will increase \citep{Lithwick2012,Batygin2013,Pichierri2019}. Thus, an initial population near exact resonance that is subjected to energy dissipation (e.g., eccentricity damping) will naturally form the trough and peak inside and outside the resonance, respectively. Several sources of dissipation have been suggested, including tides due to orbital eccentricity \citep{Delisle2014}, tides due to planetary obliquity \citep{Millholland2019}, and damping from leftover planetesimals, which can also drive divergent migration \citep{Chatterjee2015}.

We argue here that many of these models cannot explain the results of Section \ref{sec:ttv}. The fundamental issue is that they invoke strong energy dissipation to grow $\Delta$, a process that will invariably cause the systems to settle into their equilibrium with small free eccentricity and librating $\psi$. That configuration is robustly ruled out by our analysis of the TTV sample in Figures \ref{fig:delta_psi} and \ref{fig:Delta_Z}. To demonstrate this, we built four simplified population synthesis models based on published hypotheses for the \textit{Kepler} near-resonant systems, summarized in Table \ref{tab:toy_models}. Of them, stochastic forces present during disk migration best replicate the trends observed in Section \ref{sec:ttv}.

\begin{table*}[]
    \centering
    \begin{tabular}{c|c|c}
         Model & Controlling Parameter & Parameter Range \\
         \hline
         Laminar disk & Damping ratio $K\equiv \tau_a/\tau_e$ & $[10^2,10^5]$ \\
         Turbulent disk & Stochastic force strength $\kappa$ & $[10^{-7},10^{-5}]$ \\
         Tidal damping & Cumulative $e$ damping timescales & $[10^{-1},10^{2}]$ \\
         Planetesimal damping & Planetesimal disk mass $M_d$ & $[10^{-1},10^{1}]~M_\oplus$
    \end{tabular}
    \caption{An overview of our population synthesis models and the parameters that control their evolution.}
    \label{tab:toy_models}
\end{table*}

\subsection{Laminar Disk Migration}
\label{sec:disk}
Several authors \citep[e.g.][]{Choksi2020,Charalambous2022} have invoked disk migration and capture into resonance as the dominant physical processes that produce the near resonant pairs. Typically, each planet is assigned a semi-major axis and eccentricity damping timescale, $\tau_{a,i}=a_i/\dot{a}_i$, and $\tau_{e,i}=e_i/\dot{e}_i$, respectively. In the case of convergent migration with eccentricity damping (i.e. $\tau_{e,i}<0$ and $\tau_{a,1}>\tau_{a,2}$), at equilibrium the planets are wide of the resonance by
\begin{equation}
    \Delta_\text{eq} \approx 1.1~\frac{m_1+m_2}{M_*}\sqrt{k\frac{\tau_{a, \text{rel}}}{\tau_e}}
    \label{eq:Delta_eq}
\end{equation}
where $\tau_{a, \text{rel}}^{-1}\equiv\tau_{a,1}^{-1}-\tau_{a,2}^{-1}$ \citep{Terquem2019}. We have also made use of the compact orbits approximation (see Section \ref{sec:res}) and assumed $\tau_e\equiv\tau_{e,1}=\tau_{e,2}$.

Equation \ref{eq:Delta_eq} presents an immediate challenge. For our fiducial system and a distance of $\Delta=0.01$, the ratio of timescales must be $\tau_a/\tau_e\sim 10^5$. However, standard disk models predict $K\equiv\tau_a/\tau_e\sim (h/r)^{-2}\sim 100-400$, where $h/r$ is the disk aspect ratio \citep{Papaloizou2000}. Various suggestions have been made that $K$ could be higher, including flared disks \citep{Ramos2017}, torque-free inner disk edges \citep{Xiang-Gruess2015}, and alternative planet-disk interaction prescriptions \citep{Charalambous2022}. Some authors have also explicitly incorporated self-consistent disk models and torque calculations, with similar results to increasing $K$ \citep{Migaszewski2015,Cui2021}.

\begin{figure}
    \centering
    \includegraphics[width=\columnwidth]{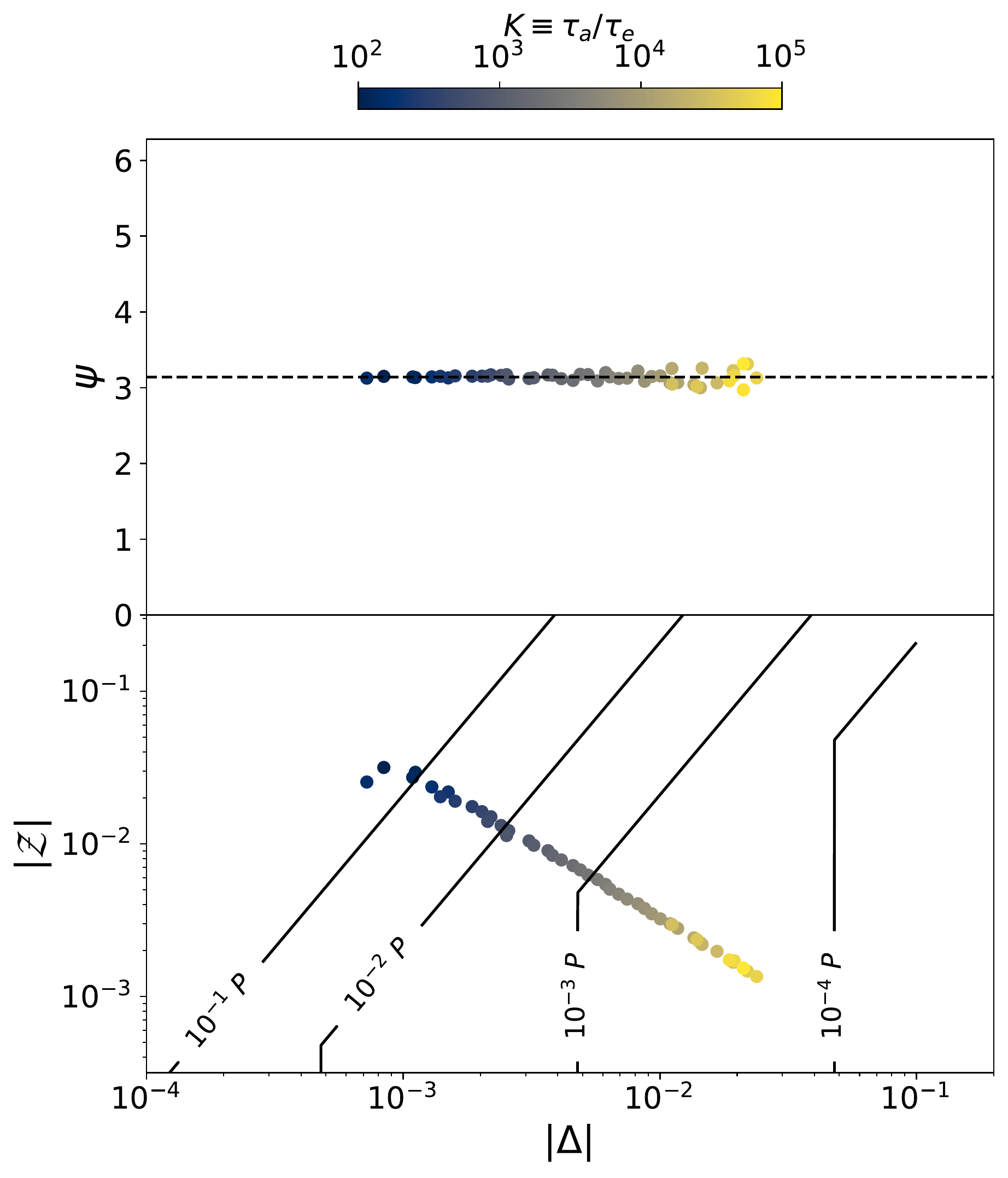}
    \caption{The results of the ``laminar disk'' population synthesis. The top panel shows $|\Delta|$ versus $\psi$, equivalent to Figure \ref{fig:delta_psi}. The bottom panel shows $|\Delta|$ versus $|\mathcal{Z}|$, equivalent to Figure \ref{fig:Delta_Z}. Points are color coded by the ratio of semi-major axis to eccentricity damping timescales $K$. Contour lines and labels show the expected amplitude of transit timing variations (Eq.~\ref{eq:TTV_amp}).}
    \label{fig:circ_disk}
\end{figure}

However, higher values of $K$ are only more efficient at damping eccentricity and finding the equilibrium. To test this, we simulated resonant capture for our fiducial system numerically. For this and all following integrations, we used the hybrid \texttt{mercurius} integrator implemented in the \texttt{rebound} package and the \texttt{reboundx} extension for disk-induced forces \citep{Rein2015,Rein2019,Tamayo2020}. We initialized the planets on circular orbits at $\Delta_\text{init}=0.05$, and set $\tau_e=-10^5$ for both planets in units of the inner orbital period. Only the outer planet was made to migrate with $\tau_a=K\cdot\tau_e$, where $K$ was varied from $10^2$ to $10^5$. When the integration time reached approximately $\tau_a/10$, the resonance was captured, the disk was removed adiabatically by increasing $\tau_a$ and $\tau_e$. The results are shown in Figure \ref{fig:circ_disk}. Although the final pairs span a large range in $\Delta$, the mixed resonant angle librates with small amplitude and hence clusters near $\psi=\pi$. Similarly, the eccentricities follow the forced eccentricity curve and depend directly on $\Delta$. Neither result is consistent with the TTV observations, which show a distinct break at $\Delta\approx 0.006$ and nonzero free eccentricities for many systems.

\subsection{Turbulent Disks}
Real protoplanetary disks are expected to have turbulent inner regions, where the magneto-rotational instability operates \citep{Nelson2004,Flock2017}. The associated density fluctuations lead to stochastic gravitational forcing on the planet not captured by simple migration and eccentricity damping timescales \citep{Nelson2004}. Stochastic forcing has been invoked to explain the smooth period ratio distribution \citep{Rein2012a}, large libration amplitudes of resonant planets \citep{Nesvorny2022}, and escape from resonance \citep{Rein2009,Batygin2017}. Turbulent fluctuations have also been shown to be consistent with planet population synthesis models that include a phase of resonant capture \citep{Izidoro2017}.

The strength of turbulent forces in real disks is highly uncertain, so here we invoke a broad range of stochasticity. We implement stochastic forces with the \texttt{reboundx} package, in which the strength of the forces is parameterized by $\kappa$, the ratio of the stochastic force to the stellar gravitational force \citep{Rein2022}. The forces themselves have an autocorrelation time equal to the planet orbital period. We ran 100 simulations of our fiducial system, with the same initial conditions as Section \ref{sec:disk} but setting $K=100$ and adding stochastic forces to both planets with a $\kappa$ that varied uniformly in log space from $10^{-7}$ to $10^{-5}$. Values of $\kappa$ can also be related to the dimensionless disk viscosity $\alpha$ via diffusion coefficients \citep{Rein2012a,Batygin2017}. Assuming a local disk surface density at $0.1$ AU of $17000$~g/cm$^2$ \citep{Batygin2017}, our range of $\kappa$ approximately corresponds to a range in $\alpha$ of $10^{-5}$ to $10^{-1}$. After $10^6$ orbits of the inner planet, the disk forces were adiabatically removed and the system was integrated for another $5\times 10^5$ inner orbits.

\begin{figure}
    \centering
    \includegraphics[width=\columnwidth]{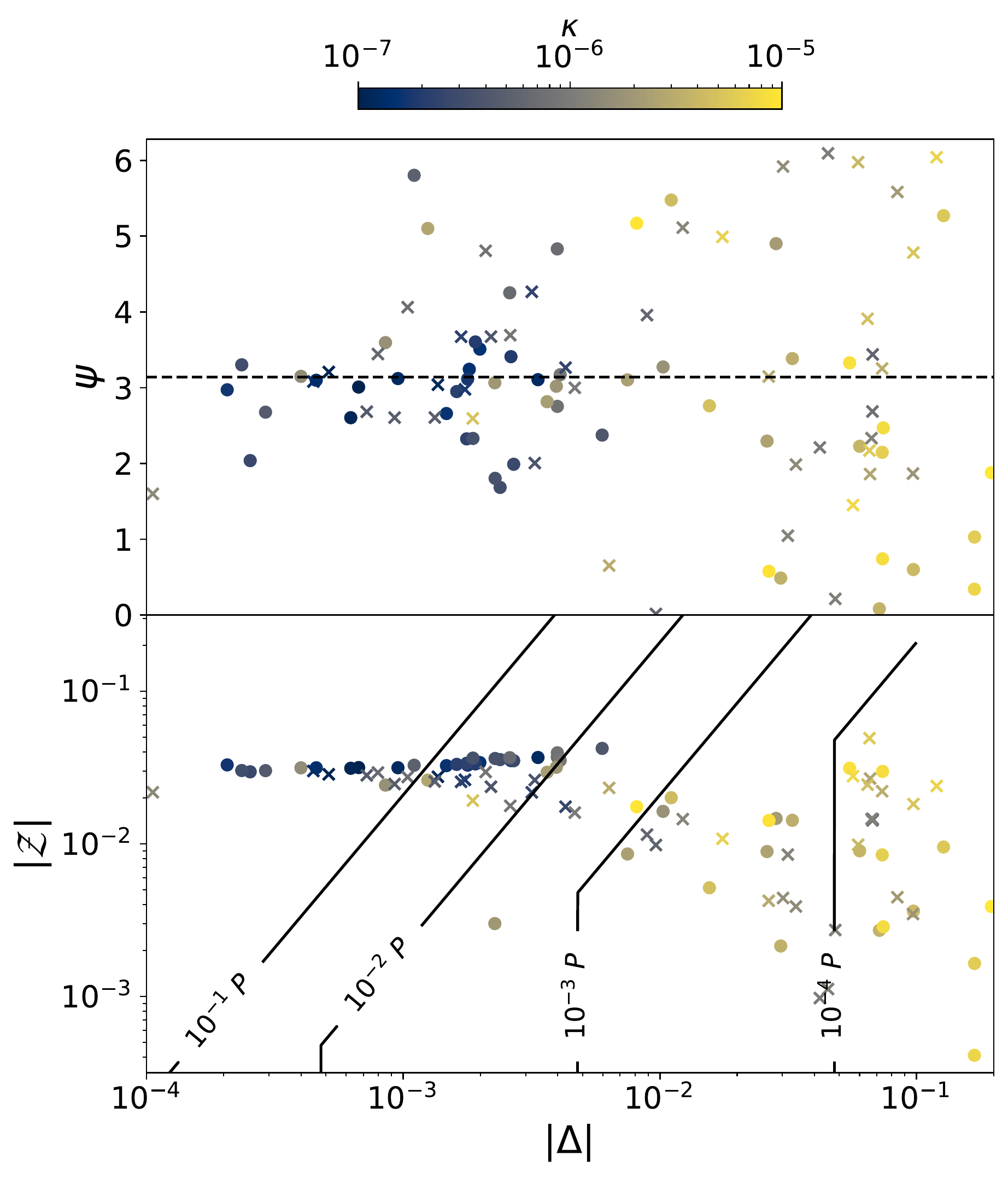}
    \caption{The same as Figure \ref{fig:circ_disk}, but for the ``turbulent disk'' model. The points are color-coded by $\kappa$, the ratio of stochastic force to stellar gravitational force.}
    \label{fig:stoch_disk}
\end{figure}
The results, plotted in Figure \ref{fig:stoch_disk}, are distinctly different than smooth disk migration (Figure \ref{fig:circ_disk}) and qualitatively similar to the observed distribution. The final distance from resonance $\Delta$ is closely related to $\kappa$. Planets that experienced a small amount of turbulence remain in the resonance but have an excited libration amplitude. Inversely, planets for which $\kappa \gtrsim 10^{-6}$ escape the resonance entirely, and by virtue of their eccentricities being stochastically driven to $\approx 0.03$, have a circulating resonant angle. We also ran another set of simulations fixing $K=300$. The results were the same as Figure \ref{fig:stoch_disk}, except that there were fewer systems with $|\Delta|<10^{-3}$ (as expected from Eq. \ref{eq:Delta_eq}). We experimented with longer integrations and found that systems with the largest values of $\kappa$ typically escaped the 3:2 resonance and continued to migrate convergently, capturing into more compact first-order resonances. Because our model only considers proximity to the 3:2 resonance, these planets reached $\Delta < -0.1$ and thus were not considered in the final analysis. However, the systems that remained near the 3:2 resonance maintained the clustering trend seen in Figure~\ref{fig:stoch_disk} even in integrations that were three times longer.

Interestingly, this model of convergent migration with stochastic forcing naturally produces a population of planets with $\Delta\approx 0.01$ without invoking very large $K$ or strong tidal dissipation. We note that the distribution of $\Delta$ in Figure \ref{fig:stoch_disk} is controlled by the distribution of $\kappa$, which is log-uniform from $10^{-7}$ to $10^{-5}$ in our simulations. Hence the observed peak near $\Delta\approx 0.01$ could be a consequence of a corresponding peak in the distribution of $\kappa$, and thus in $\alpha$. Within the formally resonant region, there appear to be too many systems with small but negative $\Delta$, although these could conceivably be moved to positive $\Delta$ via a small degree of tidal damping without disrupting the distribution of $\psi$.

\subsection{Tidal Damping}
An alternative mechanism that has been invoked in the literature to explain the deviation from resonance is a dissipative force that acts after the protoplanetary disk is gone. In contrast to the case of disk migration, there is no equilibrium: as long as the force is present, the pair of planets will continue to diverge from exact resonance \citep{Lithwick2012}. A natural source of dissipation is tides raised on the planet by the star \citep{Delisle2014}. There are major problems with this proposed solution however, including requiring anomalously small tidal quality factors \citep{Lee2013}, too high initial eccentricities \citep{Silburt2015}, and the lack of an expected signature of dependence on orbital separation \citep{Choksi2020}. Tides strengthened by planetary obliquity \citep{Millholland2019} may alleviate these issues somewhat, but not fully.

Regardless of the exact mechanism, tidal damping away from exact resonance involves considerable energy dissipation. To test the effect of this, we ran 200 integrations with $10~M_\oplus$ planets near the 3:2 resonance. The initial eccentricities were $0.01$ and $\varpi$ was drawn randomly from a uniform distribution. For each simulation, we set the eccentricity dissipation timescale for each planet to $\{10^4,10^5,10^6,10^7\}$ inner orbital periods and initial $\Delta$ uniformly from $[-0.05,0.05]$. The simulations were run for $10^6$ inner orbital periods, so that some systems experienced many damping timescales and others did not finish a single one.

\begin{figure}
    \centering
    \includegraphics[width=\columnwidth]{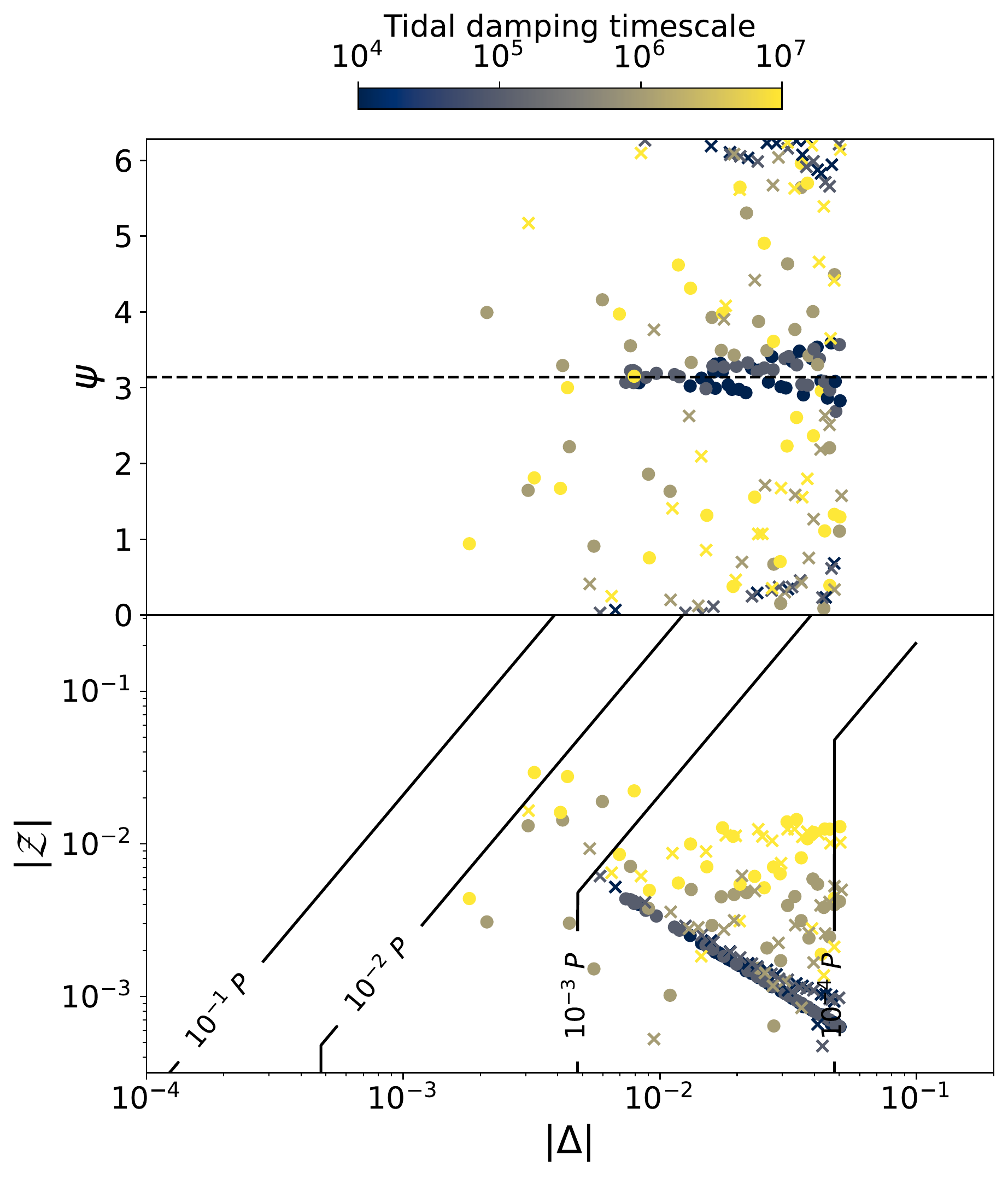}
    \caption{The same as Figure \ref{fig:circ_disk} but for the "Tidal damping" model. Points are color-coded by the timescale of eccentricity damping. In contrast to the other three models, the initial value of $\Delta$ was drawn uniformly from $[-0.05,0.05]$.}
    \label{fig:tides}
\end{figure}

The results of these integrations are shown in Figure \ref{fig:tides}. When tidal timescales are much longer than the integration duration, $\Delta$ does not change much and $\psi$ remains in circulation. Inversely, when many tidal timescales elapse, the region of very small $|\Delta|$ is cleared out and $\psi$ settles at an equilibrium value. Specifically, highly damped systems that end with $\Delta < 0$ go to the $\psi=0$ equilibrium while those that end with $\Delta > 0$ go to the $\psi=\pi$ equilibrium. Neither trend agrees with the near-uniform distribution of $\psi$ seen for the $|\Delta|>0.006$ systems in Section \ref{sec:ttv}.

\subsection{Planetesimal Interactions}
Other authors have suggested that a planetesimal disk, made of material that did not coalesce into planets, is responsible for damping and/or migration away from resonance \citep{Chatterjee2015,Ghosh2023}. To reproduce the overpopulation of planet pairs wide of resonance, they place a population of planetesimals in orbit around a resonant pair of planets, and the resulting gravitational interactions increase $\Delta$ with a dependence on the local mass of material in the planetesimal disk. While the stochastic nature of planetesimal interactions can increase the phase space area somewhat, the broad effect is to act as an energy sink and damp the planet's eccentricity through dynamical friction.

To investigate this further, we ran a set of simulations that included a planetesimal disk, similar to the setup of \cite{Chatterjee2015,Ghosh2023}. To initialize the simulations, we applied gas-driven migration and eccentricity damping timescales as in Section \ref{sec:disk} with $K=100$ and removed the gas disk adiabatically. Once the gas was completely removed, we instantaneously added a planetesimal disk of $1000$ equal-mass particles with total mass $M_d$. We varied $M_d$ across 100 simulations log-uniformly from $10^{-1}$ to $10^{1}~M_\oplus$. The planetesimals were placed randomly at radii such that their surface density scales as $\Sigma(r) \propto r^{-2}$, the approximate steady state distribution for radially-drifting dust \citep{Youdin2004,Armitage2020}. Following \cite{Ghosh2023}, we set the inner and outer edges of the disk to be the 1:3 and 3:1 resonances of the inner and outer planets, respectively. The eccentricities and inclinations (in radians) of the planetesimals were drawn uniformly from $[0,0.01]$ to match the self-consistent simulations of \cite{Chatterjee2015}. The remaining angular orbital elements were drawn uniformly from $[0,2\pi]$. These integrations were run for $2\times 10^6$ inner orbits. Planetesimals which passed within $1R_\odot$ of the central star or $2R_\oplus$ of either planet were merged with the nearby body while conserving linear momentum.

\begin{figure}
    \centering
    \includegraphics[width=\columnwidth]{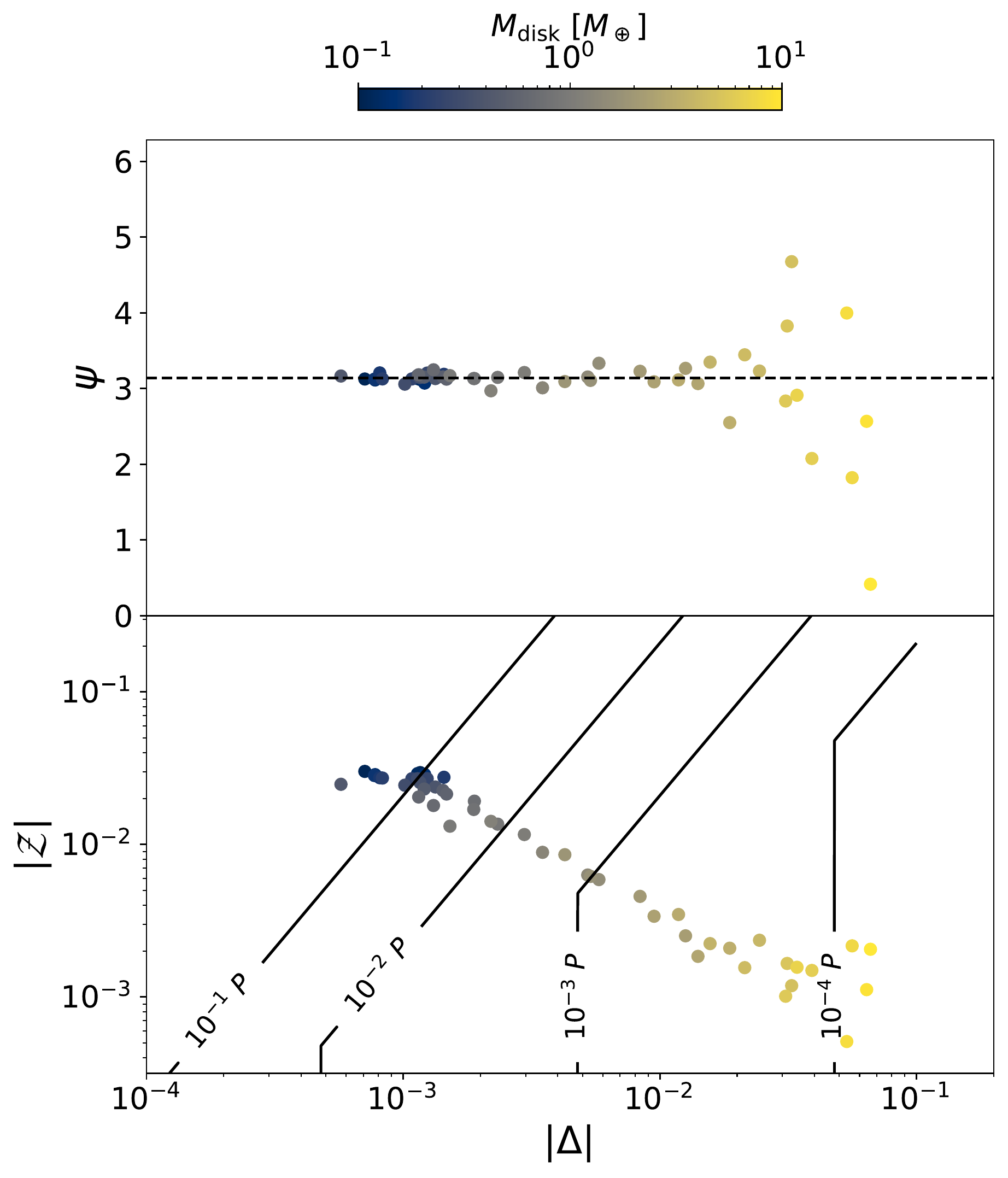}
    \caption{Same as Figure \ref{fig:circ_disk}, but for the ``Planetesimal disk'' model. The points are color-coded by $M_\text{disk}$, the total mass of planetesimals.}
    \label{fig:planetesimals}
\end{figure}

The results of this final suite of simulations are shown in Figure \ref{fig:planetesimals}. Broadly, the systems remain near the resonant equilibrium like in the laminar disk model, except for a large increase in libration amplitude for $\Delta \gtrsim 0.05$. Though qualitatively similar to the break at $\Delta \approx 0.006$ in Figure \ref{fig:delta_psi}, this break is nearly an order of magnitude more distant from exact resonance. Additionally, the final eccentricities are small and depend strongly on $\Delta$, a trend not seen in the data (Figure \ref{fig:Delta_Z}). We note that \cite{Ghosh2023} highlight the importance of a mixture model in which some systems begin not in resonance; these systems only experience limited eccentricity damping during the planetesimal migration phase and retain circulating resonant angles. Nevertheless, the dominant population within the peak around $\Delta\approx 0.01$ will be planet pairs that capture into resonance in the gas disk and will be highly damped after planetesimal interactions.

The planetesimal damping model also presents problems from the standpoint of model testing. While the planet formation process is certainly not 100\% efficient, the true quantity and distribution of unaccreted material is complex, highly uncertain, and dependent on dust and planet migration as well as detailed disk structure \citep{Hansen2012,Drazkowska2016,Raymond2020}. Furthermore, the model seems to require a degree of fine-tuning: planets that begin migrating through a disk `run away' as long as material is present \citep[see][]{Ormel2012}. When integrated long enough, many initial conditions bring planets to $\Delta\sim 0.1$ rather than the observed peak at $\Delta\approx 0.01-0.02$ \citep{Ghosh2023}.

\section{Discussion}
\subsection{Dependence on multiplicity and resonant index}
Many of the near-resonant systems within the \textit{Kepler} sample are planet pairs, and for simplicity, we have assumed in our formation models that there are only two planets in the system and that they begin near the 3:2 resonance. The canonical transformation that replaces $\phi_1$ and $\phi_2$ with $\psi$ in the first-order resonant term is not strictly valid for 3 or more planets. Additionally, the compact orbits approximation is acceptable for $k\geq 3$ but breaks down for $k=2$, approximately $1/3$ of our sample. 

To evaluate the dependence of our observational results on transit multiplicity, we split the sample into two subsamples of systems: one where only two transiting planets were detected and one where more than two were seen. Both subsamples show the libration-circulation break seen at $|\Delta|\approx 0.006$. However, there is a noticeable dearth of two-planet systems very close to commensurability. The eight systems with smallest $|\Delta|$ have three or more transiting planets, despite two-planet systems making up 41 of the 105 planet pairs. For resonant index, we performed a similar exercise by splitting the sample into two subsamples with $k=2$ and $k\geq 3$. Of the 22 systems with $|\Delta|<0.006$, only 2 are associated with the 2:1 resonance, despite that resonance accounting for 40 of the 105 planet pairs. A possible explanation for these trends is that low disk turbulence, relative to the migration rate, skips capture into the 2:1 resonance and delivers systems deep (i.e. small $|\Delta|$) into higher-index resonances. On the other hand, high turbulence may disrupt the most compact systems, leaving only circulating systems near the 2:1 resonance. We encourage more work on this topic.

\subsection{Limitations of our work}
Our analysis does not take into account the effects of sampling bias. Importantly, \cite{Hadden2017} and \cite{Jontof-Hutter2021} do not model systems with weak or undetectable TTVs. This choice preferentially discards lower-mass planets, pairs more distant from resonance, and pairs with small free eccentricity \citep{Hadden2016}. Our results should be robust to the first two effects because we do not directly model the distributions of planetary masses or $\Delta$. The final effect suggests that some nearly circular systems might be missing from Figure \ref{fig:Delta_Z} at higher $\Delta$. However, Eq. \ref{eq:TTV_amp} indicates that the dependence on eccentricity is only important for $|\mathcal{Z}| \gtrsim |\Delta|$. Thus, highly damped systems at $\Delta \sim 0.01$ would be observable if they existed, but they are not seen in Figure \ref{fig:Delta_Z}.

An additional sytematic bias in the models of \cite{Hadden2017} and \cite{Jontof-Hutter2021} is that they only consider the TTV contributions from known, transiting planets. Unseen planets may induce a TTV signal that is interpreted as coming from one of the transiting planets, biasing the measured $\psi$ and $|\mathcal{Z}|$. 

\subsection{Eccentricity Excitation or Damping?}
In general, it is not obvious what sets the eccentricities of planets in multiplanet systems. Transit timing and transit duration measurements have independently agreed that typical eccentricities in multi-planet systems are small but nonzero \citep{Hadden2017,VanEylen2019}. There is no evidence of correlation between eccentricity and any system properties except multiplicity \citep{VanEylen2019,He2020}. Eccentricities must be bounded below by self-excitation (even for initially circular orbits) and above by orbit-crossing and stability limits. Remarkably, the observed census of planetary systems falls in between these two regimes: planets are neither dynamically cold, nor do they reside right at the stability boundary \citep{Yee2021}. Therefore, planet formation scenarios that rely on strong damping must eventually include a source of eccentricity excitation; alternatively, scenarios that invoke dynamical sculpting as the dominant process require damping. Ultimately, the full story of planet formation must incorporate a competition between mechanisms of eccentricity damping and excitation.

\section{Conclusion}
In this work, we have reanalyzed the near-resonant planetary systems characterized with transit timing variations from \textit{Kepler}. We show that despite fundamental limitations in TTV interpretation, the resonant behavior of these systems can be probed in detail. Planet pairs very close to exact resonance ($|\Delta|<0.006$) have a librating mixed resonant angle, but those in the peak $\sim 1\%$ wide of resonance are predominantly circulating. This result is difficult to reconcile with several hypotheses which argue that dissipative processes place pairs of planets wide of resonance, keeping them in a stable quasi-equilibrium state. Stochastic forces during migration, meant to simulate density variations in a turbulent gaseous disk, offer a promising explanation for the qualitative features of the sample.

Future work should use a more thorough modeling effort that consider a mixture of resonant and non-resonant systems. Recent theories of planet formation have argued that near-universal dynamical instabilities successfully reproduce the observed mostly smooth period ratio distribution \citep{Izidoro2017}. In such a model, some planet pairs are `near-resonant' only coincidentally (reaching that state after the dissipation of the protoplanetary disk) and not as the consequence of resonant capture or damping. The actual fraction of systems that never experienced a post-gas instability in the overall sample is unclear: \cite{Izidoro2021} suggest that no more than $5\%$ remain stable and resonant, but it remains to be seen whether those results are consistent with the overabundance of three-body libration seen in the \textit{Kepler} sample \citep{Goldberg2021,Cerioni2022}. While basic modeling of the peaks in the period ratio distribution has had some success \citep[e.g.][]{Choksi2020,Ghosh2023}, the results of this work illuminate a novel and stringent constraint that must be accounted for in a complete model of planet formation.

\begin{acknowledgements}
We are grateful to the anonymous referee for a thorough reading and useful recommendations that substantially improved this work. We thank Jon Zink and Juliette Becker for insightful suggestions. While this work was in peer review, we became aware that \cite{Choksi2023} also arrived at some of the same results presented in this work simultaneously and independently. K. B. is grateful to Caltech, the Caltech Center for Comparative Planetary Evolution, the David and Lucile Packard Foundation, and the Alfred P. Sloan Foundation for their generous support.
\end{acknowledgements}

\bibliography{main}
\bibliographystyle{aasjournal}

\end{document}